\documentclass[%
11pt,aps,letterpaper,
typeset,onecolumn,superscriptaddress,
showpacs,floatfix]{quantumarticle}
\pdfoutput=1
\usepackage{dcolumn}
\usepackage{bm}
\usepackage{graphicx} 
\usepackage{amsmath}
\usepackage{amssymb}
\usepackage{physics}
\usepackage{braket}
\usepackage{hyperref}

\DeclareMathOperator*{\argmin}{arg\,min}
\newtheorem{theorem}{Theorem}
\newtheorem{proposition}{Proposition}
\newtheorem{corollary}{Corollary}
\usepackage{subcaption}
\usepackage{setspace}
\begin{document}

\title{Rigorous quantum state tomography for distributed quantum computing}
\author{Hans M\"attig-V\'asquez}
\email{hmattig@udec.cl}
\affiliation{Instituto Milenio de Investigaci\'on en \'Optica y Departamento de F\'isica, Facultad de Ciencias F\'isicas y Matem\'aticas, Universidad de Concepci\'on, Concepci\'on, Chile}
\author{Aldo Delgado}
\affiliation{Instituto Milenio de Investigaci\'on en \'Optica y Departamento de F\'isica, Facultad de Ciencias F\'isicas y Matem\'aticas, Universidad de Concepci\'on, Concepci\'on, Chile}
\author{Luciano Pereira}
\email{luciano.pereira@icfo.eu}
\affiliation{ICFO - Institut de Ciencies Fotoniques, The Barcelona Institute of Science and Technology, 08860 Castelldefels, Barcelona, Spain}

\begin{abstract}
    Distributed quantum computing offers a promising approach to scaling quantum devices by networking multiple quantum processors. We present a quantum state tomography protocol tailored for distributed quantum computers that avoids assuming remote entanglement as a primitive resource. The protocol extends projected least-squares (PLS) tomography based on projective 2-designs to systems composed of multiple quantum processors, using only local operations within each processor and classical communication between nodes. Assuming entanglement within each individual quantum processor is trusted, the protocol can be executed using mutually unbiased bases. We derive rigorous, non-asymptotic trace-norm error bounds for the PLS estimator, with explicit exponential dependence on the number of nodes. In addition, we establish certified error bounds for estimating entanglement negativity from the PLS estimator. Numerical simulations for systems of up to seven qubits distributed across several devices validate the theoretical error bounds.
\end{abstract}

\maketitle

\section{Introduction}
Quantum computing has the potential to solve particular tasks more efficiently than classical computing \cite{Nielsen_Chuang_2010}. In recent years, significant hardware advancements have been made, including increases in the number of qubits and reductions in error rates, moving us closer to fault-tolerant quantum computing \cite{quant-ph/9605011, Grassl2009, 0904.2557}. However, scalability remains a major challenge \cite{Acharya2023_google}. As systems grow in size, errors from decoherence, control imperfections, and crosstalk accumulate, while dense integration introduces severe engineering constraints \cite{Brecht2016}. Even with quantum error correction \cite{Krinner2022,Roffe2023}, building a large-scale individual quantum computer is very demanding, leading to the exploration of alternative scalable architectures.

Distributed quantum computing addresses scalability challenges by connecting multiple smaller processors via classical or quantum communication \cite{Caleffi2024,Barral2025,2510.15630}. By networking independent quantum computers, it is possible to effectively expand the computational space while enabling modular architectures in which each node can be fabricated, calibrated, and upgraded independently. Distributed quantum computing has been experimentally demonstrated at small scale in systems such as superconducting circuits \cite{PhysRevLett.125.260502} and trapped ions \cite{Main2025}, including the generation of remote entangled states \cite{Storz2023} and the execution of distributed quantum algorithms \cite{Yimsiriwattana2004,Main2025}.

In this context, characterization algorithms tailored to distributed quantum computers are essential \cite{Storz2025_ST,Dalton2025_CPV}, particularly to characterize remote entanglement across multiple quantum nodes. Quantum tomography \cite{Paris2004,Banaszek2013} provides a natural framework for this task. The most standard and experimentally friendly approach to tomography is using Pauli measurements, which, with only local operations, provides an estimator with error $\epsilon$ with sample complexity $\mathcal{O}(d^{1.6}r^2\log d/\epsilon^2)$. Projective 2-designs, such as mutually unbiased bases \cite{Klappenecker2005} or symmetric informationally complete measurements \cite{Renes2004}, can also be used for tomography. They provide a better sample complexity of $\mathcal{O}(dr^2\log d/\epsilon^2)$, but require an arbitrary entangled operation for its implementation. 
Consequently, they may require remote entanglement in a distributed quantum computer, leading to a potentially inconsistent characterization since the resource being characterized is already assumed within the protocol itself.

In this work, we introduce a projected least-squares (PLS) state tomography tailored for distributed quantum computers. We suppose that entanglement within each processor is trusted, but that remote entanglement is noisy and untrusted. Our protocol applies not only to future distributed fault-tolerant quantum computing \cite{Katabarwa2024_EFT,Larasati2025} but also to near-term devices by splitting a single large device into several virtual high-quality nodes. The protocol is based on performing measurements in the tensor product of projective 2-designs at each node, requiring only local operations within each processor and avoiding remote entanglement. The resulting empirical frequencies, obtained via classical communication, are used to construct a least-squares estimator of the unknown state that admits a closed-form expression due to the node 2-design structure of the measurements. Since this estimator is not guaranteed to be physical, it is projected onto the set of positive semidefinite matrices, yielding the PLS estimator.  

We derive rigorous, non-asymptotic sample complexity bounds for the PLS estimator using concentration inequalities for random matrices \cite{Tropp2011, Guta2020}. To obtain an estimate with trace-norm error $\epsilon$, our protocol requires a sample of size $\mathcal{O}(2^Mr^2d\log(d)/\epsilon^2)$, which connects the tomographies via Pauli measurements and global 2-designs. The exponential dependence on the number of subsystems reflects the statistical cost of performing tomography using local 2-design measurements. In addition, we consider estimating remote entanglement using a tomographic estimator, establishing certified error bounds for the entanglement Negativity \cite{PhysRevA.65.032314}. These bounds generally apply to any tomographic estimator satisfying a trace-norm error bound, and, in particular, to the PLS estimator obtained from the proposed protocol.

We validate the theoretical results through numerical simulations of reconstructing Haar-random states with 2 to 7 qubits. The 2-design measurements at each node are efficiently implemented using mutually unbiased bases \cite{Ivonovic1981, Wootters1989}. We compare tomography performed within an individual quantum computer to that performed across several distributed quantum computers, considering all possible qubit splittings. We observe that the reconstruction error in distributed systems is larger than in individual systems, in agreement with the theoretical error bounds.

\section{Tomographic Protocol}

\begin{figure*}[t!]
    \centering
        \centering
        \includegraphics[width=\linewidth]{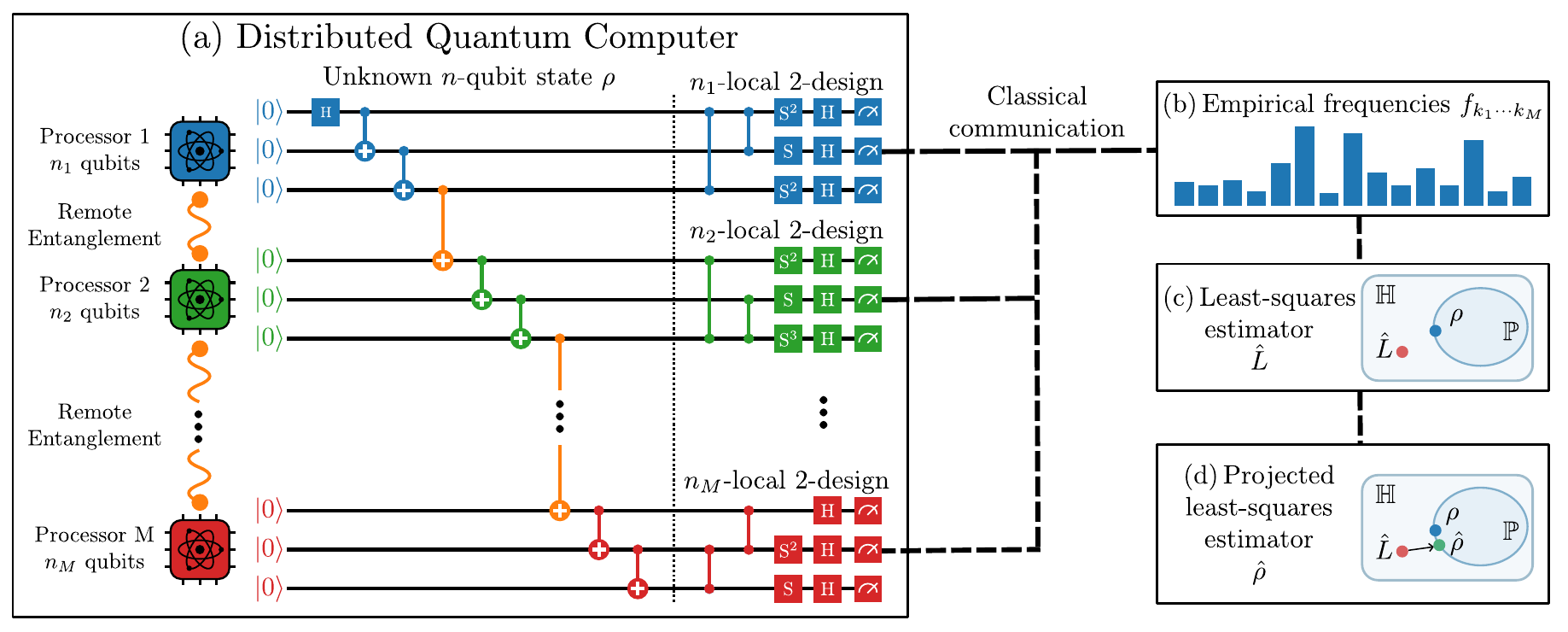}
    \caption{ Diagram of the PLS tomography in a distributed quantum device. (a) The circuit illustrates a quantum circuit for implementing tomography across several distributed devices (blue, green, red) connected by remote entanglement (orange). Initially, we prepare the unknown state; in this example, we use a multi-qubit GHZ state. This state preparation requires both node-local operations (blue, green, red)  and remote entanglement (orange). Thereafter, measurements in node 2-designs are carried out at each quantum node, requiring only node-local operations (blue, green, red). Steps for the PLS tomography: (b) empirical frequency acquisition, (c) least-squares estimation, and (d) projection.}\label{fig:diagram}
\end{figure*}

Consider an $n$-qubit quantum system composed of $M$ subsystems, where the $j$-th subsystem consists of $n_j$ qubits. Each subsystem, of dimension $d_j = 2^{n_j}$, represents a node of the distributed quantum computer, as is displayed in Fig.\thinspace\ref{fig:diagram}(a). Let 
\begin{equation}
    \mathcal{D}_j = \{ \dyad*{v_k^j} \}_{k=1}^{m_j}
\end{equation}
be a projective 2-design with $m_j$ elements for the $j$-th subsystem. This means that they reproduce the second-order moments of the Haar distribution   
\begin{align}
    \frac{1}{m_j} \sum_{k=1}^{m_j} (\dyad*{v_k^j})^{\otimes 2} = \int_{\mathbb{S}^{d_j}}\thinspace{\rm d}\mu(v)  (\dyad{v})^{\otimes 2}, 
\end{align}
with $\mathbb{S}^{d_j}$ the unit sphere in $\mathbb{C}^{d_j}$ and ${\rm d}\mu(v)$ the unitary invariant Haar measure \cite{Mele2024introductiontohaar}. Then, the set of $n_j$-local 2-designs given by
\begin{align}
    \mathcal{M} = \left\{ \frac{d}{m} \bigotimes_{j=1}^M \dyad*{v_{k_j}^j} \right\}_{k_j=1}^{m_j}, \label{eq:POVM_local}
\end{align}
with $d = \prod_j d_j$ and $m = \prod_j m_j$, defines a positive operator-valued measure (POVM) in the composite system. We refer to this measurement as a node 2-design. Importantly, this POVM is fully separable across subsystems and therefore does not require remote entanglement for its implementation. In the particular case $M=n$, each node contains a single qubit, and the POVM reduces to fully local single-qubit measurements. POVM $\mathcal{M}$ can be realized using mutually unbiased bases (MUBs) at each node. These bases provide a minimal measurement strategy for reconstructing unknown multiqubit states. They can be implemented with the algorithm introduced in \cite{2311.11698}, which generates quantum circuits with $\mathcal{O}(n^2)$ gates for each of the $2^n+1$ MUBs in a computational time complexity of $\mathcal{O}(n^3)$. Despite polynomial scaling, their realization remains challenging due to the need for entangling operations. In this scenario, we said that an entangling gate is trusted if it is implemented with an error sufficiently small such that the overall error in realizing a node 2-design is negligible. 

Suppose that the system is prepared in an unknown state $\rho \in \mathbb{H}_d$, with $\mathbb{H}_d$ the set of $d$-dimensional hermitian matrices. The tomographic protocol proceeds as follows (see Fig.\thinspace\ref{fig:diagram}(b)):
\begin{enumerate}
    \item We perform measurements corresponding to the POVM $\mathcal{M}$. This measurement can be interpreted as a linear map $\mathcal{M} : \mathbb{H}_d \rightarrow \mathbb{R}^m$ defined by 
    \begin{align} 
        X \in \mathbb{H}_d \;\longmapsto\; \mathcal{M}(X)_{k_1\cdots k_M} = \frac{d}{m}\Tr\!\left( \bigotimes_{j=1}^M \dyad*{v_{k_j}^j} \, X \right). \label{eq:linear_map}
    \end{align} 
    Performing the measurement on a sample of $N$ identically independent prepared copies of $\rho$ yields empirical frequencies $f_{k_1\cdots k_M}$, which serve as estimators of $\mathcal{M}(\rho)_{k_1\cdots k_M}$. Since each measurement operates locally within each subsystem, constructing $f_{k_1\cdots k_M}$ only requires classical communication among the nodes. 
    
    \item We construct the least-squares (LS) estimator by solving the least-squares problem
    \begin{align}
        \hat{L} = \argmin_{X \in \mathbb{H}_d} \sum_{k_1\cdots k_M = 1}^{m_1 \dots m_M}
        \left[ f_{k_1\cdots k_M} - \mathcal{M}(X)_{k_1\cdots k_M} \right]^2. \label{eq:LSP}
    \end{align}
    The solution is given by $\hat{L} = (\mathcal{M}^\dagger \mathcal{M})^{-1} \mathcal{M}^\dagger (f_{k_1\cdots k_M})$. For projective 2-designs, this estimator admits a closed-form expression,
    \begin{align}
        \hat{L} = \sum_{k_1,\cdots,k_M}^{m_1\cdots m_M} f_{k_1,\dots,k_M}\bigotimes_{j=1}^M(d_j+1)\left[\dyad*{v^j_{k_j}}{v^j_{k_j}}-\frac{\mathbb{I}_j}{d_j+1}\right], \label{eq:LSE}
    \end{align}
    with $\mathbb{I}$ the identity on the $j$-th subsystem. The proof of this can be found in  Appendix \ref{ap:SLE}.
    
    \item Since the LS estimator $\hat{L}$ is not guaranteed to be positive semidefinite, we project it onto the set of positive semidefinite matrices with unit trace, denoted by $\mathbb{P}$, by solving
    \begin{align}
        \hat{\rho} = \argmin_{\sigma \in \mathbb{P}} \| \hat{L} - \sigma \|_2 ,
    \end{align}
    with $\norm{A}_2$ the Hilbert-Schimidt norm. The solution to this problem can be obtained using an iterative algorithm introduced in~\cite{PhysRevLett.108.070502} or by semidefinite programming \cite{SDP_review}. The estimator $\hat{\rho}$ is called the projected least-squares (PLS) estimator \cite{Guta2020, SurawyStepney2022,2507.04500}.
\end{enumerate}

\section{Rigorous error bound}
Using concentration inequalities for random matrices, we derive rigorous, non-asymptotic error bounds for the PLS estimator, expressed in the following theorem: 

\begin{theorem}\label{theo:main}
    Let $\rho\in\mathbb{H}$ be a quantum state and let $\hat\rho\in\mathbb{H}$ be its PLS estimator obtained from $N$ measurements of the POVM $\mathcal{M}$. Then, for any $\epsilon\leq 1$, we have
    \begin{align}
        \mathbb{P}\left[ \norm{ \rho - \hat{\rho} }_1 \geq \epsilon \right] \leq d \exp\left(  -\frac{3}{128}\frac{\epsilon^2N}{2^M r^2 \tilde{d} } \right), 
    \end{align}
    where $r=\min\{ {\rm rank}(\rho), {\rm rank}(\hat\rho) \}$ and $\tilde{d}=\prod_{j=1}^M(d_j-1/2)$.
\end{theorem}
The demonstration is provided in Appendix \ref{ap:conc_ineq}. The following corollary establishes rigorous error bounds for the PLS estimator:
\begin{corollary}
    Following the same hypothesis as in Theorem \ref{theo:main}, we have that with probability $1-\delta$ the error in trace-norm error of the PLS estimator $\hat{\rho}$ is upper-bounded by
    \begin{align}
        \norm{\rho-\hat{\rho}}_1^2 \leq \frac{128}{3}\frac{2^M r^2 \tilde{d}}{N}\log\left(\frac{d}{\delta} \right) \label{eq:error_bound}
    \end{align}
\end{corollary}
We can see that the reconstruction quality scales exponentially with the number of subsystems $M$. This implies that treating the complete quantum system as a single device and performing tomography with global 2-designs yields the best estimation accuracy, scaling as  $\mathcal{O}(d)$. If we restrict to node 2-designs, the error bound is worse, scaling as $\mathcal{O}(2^Md)$, but it is better suited to noisy devices as we avoid using remote entanglement. The extreme case employs full local tomography via Pauli measurements, that is, $d_j=2$, but this case yields the largest error bound, scaling as $\mathcal{O}(3^M)$. Therefore, if trusted entanglement is available, it is recommended to perform tomography using the node 2-designs. 

Analogously, we can establish a corollary on the sample complexity:
\begin{corollary}
    Following the same hypothesis as in Theorem \ref{theo:main}, we have that to obtain a PLS estimation with an error of at least \(\epsilon\) with a probability of \(1 - \delta\), we need to use a sample size $N$ such that
    \begin{align}
        N \geq\frac{128}{3}\frac{2^M r^2 \tilde{d} }{\epsilon^2}\log\left(\frac{d}{\delta} \right).\label{eq:sample_complexity}
    \end{align}
\end{corollary}

Notice that in order to obtain the $\mathcal{O}(3^M)$ scaling with Pauli measurements, expressing the concentration inequality in terms of the effective dimension $\tilde{d}$ is mandatory. However, for simplicity, we will use the fact that $\tilde{d}\leq d$ and refer to the error bound and the sample complexity in terms of $d$ as $\mathcal{O}(2^Mr^2d\log(d)/N)$ and $\mathcal{O}(2^Mr^2d\log(d)/\epsilon^2)$, respectively.

\section{Entanglement estimation}

The PLS estimator $\hat{\rho}$ can be used to quantify remote entanglement. To this end, let us consider a bipartition of the system in which each partition contains $M_j$ nodes and has dimension $d_j$. A well-known entanglement measure is Negativity \cite{PhysRevA.65.032314}, defined as
\begin{align}
    \mathcal{N}(\rho) = \frac{\norm{\rho^{\top_2}}_1-1}{2}, \label{eq:negativity}
\end{align}
where $\rho^{\top_j}$ is the partial transpose of $\rho$ with respect to partition $j$. We can establish a rigorous error bound of the estimated negativity calculated from the PLS estimator. 

\begin{theorem}
Let $\rho\in\mathbb{H}$ be a quantum state and let $\hat\rho\in\mathbb{H}$ be a tomographic estimator such as $\norm{\rho -\hat\rho}_1\leq \epsilon$. Then, the error of estimating the Negativity defined by Eq.\thinspace\eqref{eq:negativity} from $\hat\rho$ is upper-bounded by
\begin{align}
    \abs{ \mathcal{N}(\rho)-\mathcal{N}(\hat\rho) } \leq \frac{\epsilon}{2}. \label{eq:bound_negativity}
\end{align}
    
\end{theorem}
The proof of this theorem is as follows: Employing the reversed triangular inequality, we have that
\begin{align}
    \abs{ \mathcal{N}(\rho)- \mathcal{N}(\hat\rho) } = \frac{1}{2}\abs{ \norm*{\rho^{\top_2}}_1 - \norm*{\hat\rho^{\top_2}}_1  } \leq \frac{1}{2}\norm{ (\rho-\hat\rho)^{\top_2} }_1.
\end{align}
Using the fact that the partial transpose preserves the trace norm and the assumption $\norm{\rho -\hat\rho}_1 \leq \epsilon$, we obtain Eq.\thinspace\eqref{eq:bound_negativity}.

This result is independent of the specific tomography protocol and applies to any estimator that satisfies a trace-norm error bound. In the specific case of the PLS from Eq.\thinspace\eqref{eq:error_bound}, we find that
\begin{align}
    \abs{ \mathcal{N}(\rho)- \mathcal{N}(\hat\rho) }^2 \leq \frac{32r^2 2^M\tilde{d}}{3N}\log\left(\frac{d}{\delta} \right),
\end{align}
with probability $1-\delta$. 

\section{Numerical Simulations}

To validate our method, we performed numerical simulations of tomographic reconstruction using Qiskit \cite{qiskit2024}. The code for the simulations is available at \cite{github}. We generated a set of 100 Haar-random pure states for systems ranging from 2 to 7 qubits \cite{Mezzadri2007,Zyczkowski2011}. This choice provides an ensemble that approximates the average reconstruction error over the Hilbert space. For each system size, we simulated tomography under two scenarios: all qubits within a single quantum device, and qubits distributed across multiple quantum nodes. To implement a node 2-design, we employed MUBs following the algorithm introduced in \cite{2311.11698}, which enables implementation using $\mathcal{O}(n^2)$ gates. The tomographic reconstruction was simulated using $N \in \{5 \times 10^6, 10^7, 1.5 \times 10^7\}$ independent and identically prepared copies of each state. To quantify the average reconstruction quality, we fitted an empirical scaling law for the average trace-norm error using the ansatz  $\bar{\epsilon}^2 = \alpha (2^M)^\beta \tilde{d}^\gamma\log(d) / N^\delta,$ over the full dataset spanning all Hilbert spaces, sample sizes $N$, and distribution configurations $M$. Note that, since the states are pure, we simply set $r=1$. From this fit, we obtained the following heuristic scaling,
\begin{equation}
\bar{\epsilon}^2 \approx 4 \frac{(2^M)^{0.8} \tilde{d}^{0.95} \log(d)}{N}. \label{eq:heuristic_scaling}
\end{equation}
The error in the fitting parameter can be found in Appendix \ref{apex:sims}. This empirical scaling is consistent with the theoretical upper bound of Eq.\thinspace\eqref{eq:error_bound} for several reasons. First, the proportionality constant is significantly smaller than that of the bound. Second, the dependence on the sample size $N$ matches the theoretical prediction. Finally, the exponents associated with $2^M$ and $\tilde{d}$ are lower, indicating improved scaling with system size. This reflects the conservative nature of the theoretical upper bounds, so that the average quality of the reconstruction is better. 

\begin{figure*}[t!]
    \centering
    \includegraphics[width=.8\linewidth]{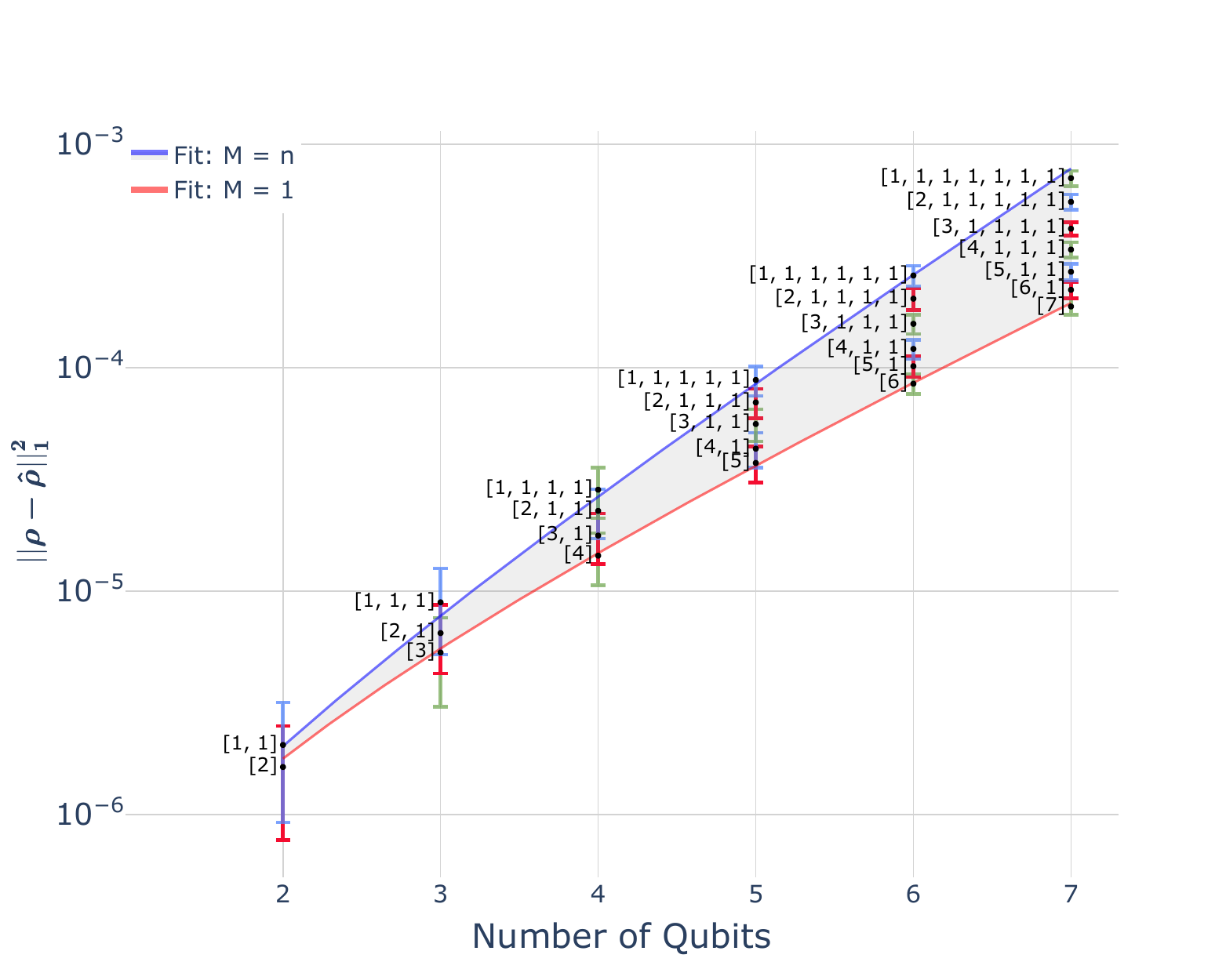}
    \caption{ Numerical simulation of PLS tomography for a sample size of $N=1.5\times10^7$. The labels next to each data point show the distribution of qubits among quantum nodes. Each point corresponds to the average trace-norm error from the tomography of 100 random Haar states, while the error bars correspond to their standard deviation. The solid lines correspond to the heuristic scaling of Eq.\thinspace\eqref{eq:heuristic_scaling} for $M = 1$ (red) and $M = n$ (blue).}\label{fig:results}
\end{figure*}

Figure~\ref{fig:results} shows the average trace-norm error as a function of the system dimension for $N = 1.5\times10^7$. Results for other sample sizes are provided in Appendix~\ref{apex:sims}. The labels next to each data point indicate the distribution of qubits across quantum nodes. Thus, the number of subsystems $M$ increases from bottom to top. More configurations than those shown in the figure were simulated to calculate heuristic scaling. However, for display purposes, these were omitted from the figure. The solid lines correspond to the heuristic scaling for the extreme cases $M = 1$ (red) and $M = n$ (blue). As predicted by the theoretical upper bound, the reconstruction quality degrades as the number of quantum nodes $M$ increases. The best performance is achieved with a global 2-design ($M = 1$), while the worst is achieved with fully local Pauli tomography ($M = n$).

\begin{figure*}[t!]
    \centering
    \includegraphics[width=.8\linewidth]{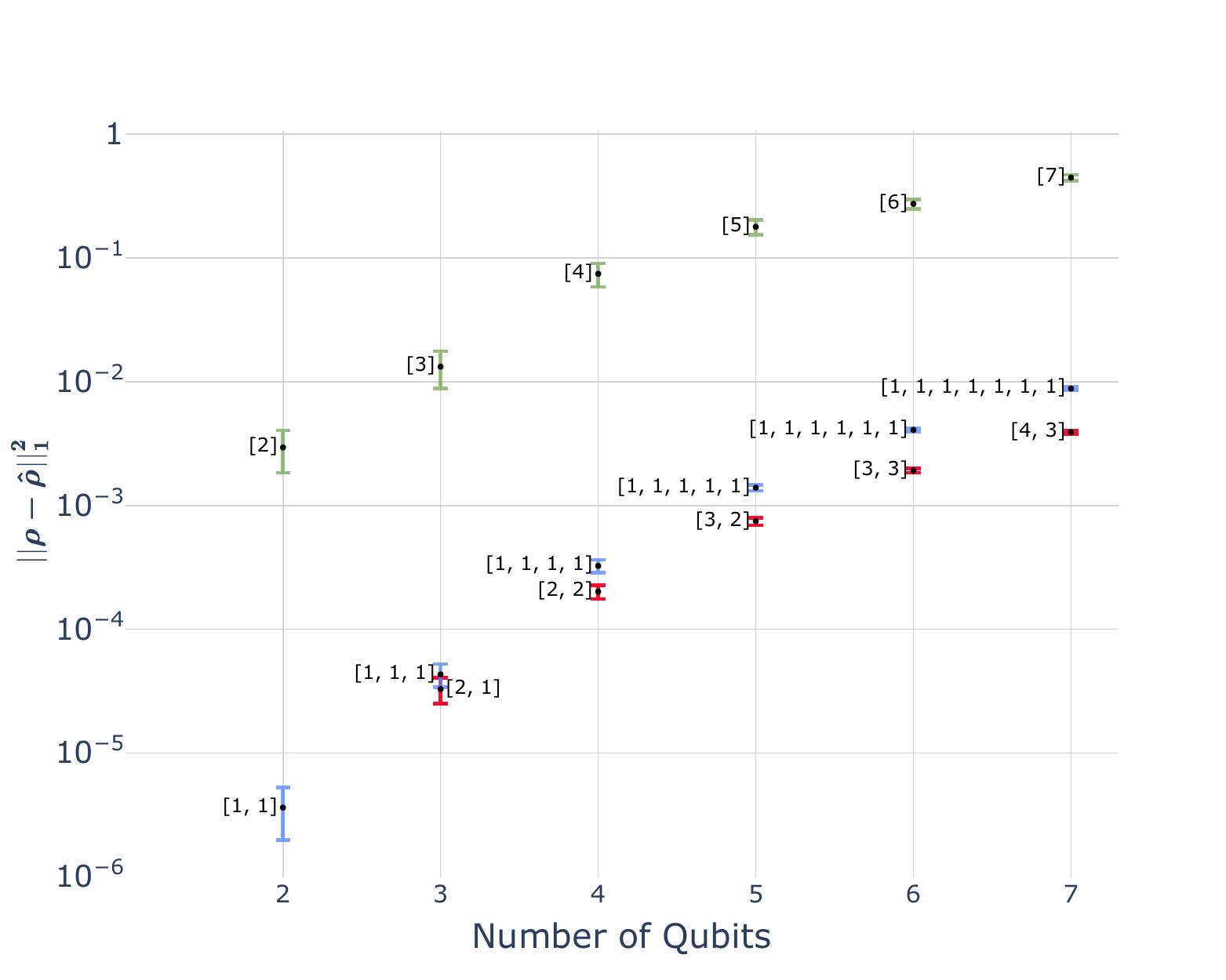}
    \caption{ Noisy numerical simulation of PLS tomography for a sample size of $N=1.5\times10^7$. The labels next to each data point show the distribution of qubits among quantum nodes. Each point corresponds to the average trace-norm error from the tomography of a 100 locally-random GHZ state, while the error bars correspond to their standard deviation. }\label{fig:results_noise}
\end{figure*}

To test the protocol in a scenario with untrusted remote entanglement, we carried out numerical simulations that include noise. We consider a quantum system composed of two quantum nodes, each with all-to-all connectivity and trusted operations. We assume that these nodes are connected by a single remote CNOT between two fixed qubits and that it is subject to depolarizing noise $\lambda = 0.05$. Numerical simulations for other noises are included in the Appendix \ref{apex:sims}. Initially, we prepared the system in a Greenberger–Horne–Zeilinger (GHZ) state, which is affected by the noisy remote CNOT gate. In appendix \ref{apx:GHZ} we show that the noisy GHZ state has rank 8. We then apply independent Haar-random unitaries at each qubit. These local random operations preserve both the entanglement structure of the GHZ state and the noise induced by the quantum link, while removing basis-dependent features. In this way, locally-random GHZ states form an ensemble that, although it does not sample the full Hilbert space, effectively isolates the effect of noisy remote entanglement. Moreover, GHZ states provide a natural and experimentally relevant benchmark for large-scale \cite{Bao2024} and distributed architectures \cite{Zhang2026}, as they maximize nonlocal correlations while requiring a minimal number of inter-node entangling operations. Thereafter, we consider reconstructing these states using local Pauli measurements, a global 2-design, and a node 2-design, with a sample size of $N=1.5\times10^7$. The global 2-design is the only approach affected by the noise from the remote CNOT gate. 

Figure\thinspace\ref{fig:results_noise} shows the average trace-norm error for the tomography of 100 the noisy locally-random GHZ state. The results indicate that tomography with the global 2-design (green) is strongly affected by noise, which prevents achieving the statistical scaling of the theoretical bound in Eq.\thinspace \eqref{eq:error_bound}. This measurement yields an error of nearly $10^{-2}$ for all inspected dimensions. In contrast, tomographies with local Pauli measurements (blue) and a node 2-design (red) were unaffected by this noise, enabling them to reliably estimate the noisy GHZ state and achieve the theoretical bound. In addition, we observe that the node 2-design achieves higher estimation accuracy and that its difference with local Pauli tomography increases with the number of qubits, highlighting its advantage for characterizing distributed devices in noisy environments. Note that the errors associated with estimating noisy states, as shown in Fig.\thinspace\ref{fig:results_noise}, are worse than those for estimating pure states in Fig.\thinspace\ref{fig:results}. This is because, according to the theoretical upper bound of Eq.\thinspace\eqref{eq:error_bound}, states with higher rank are estimated with lower accuracy.

\begin{figure*}[t!]
    \centering
    \includegraphics[width=.8\linewidth]{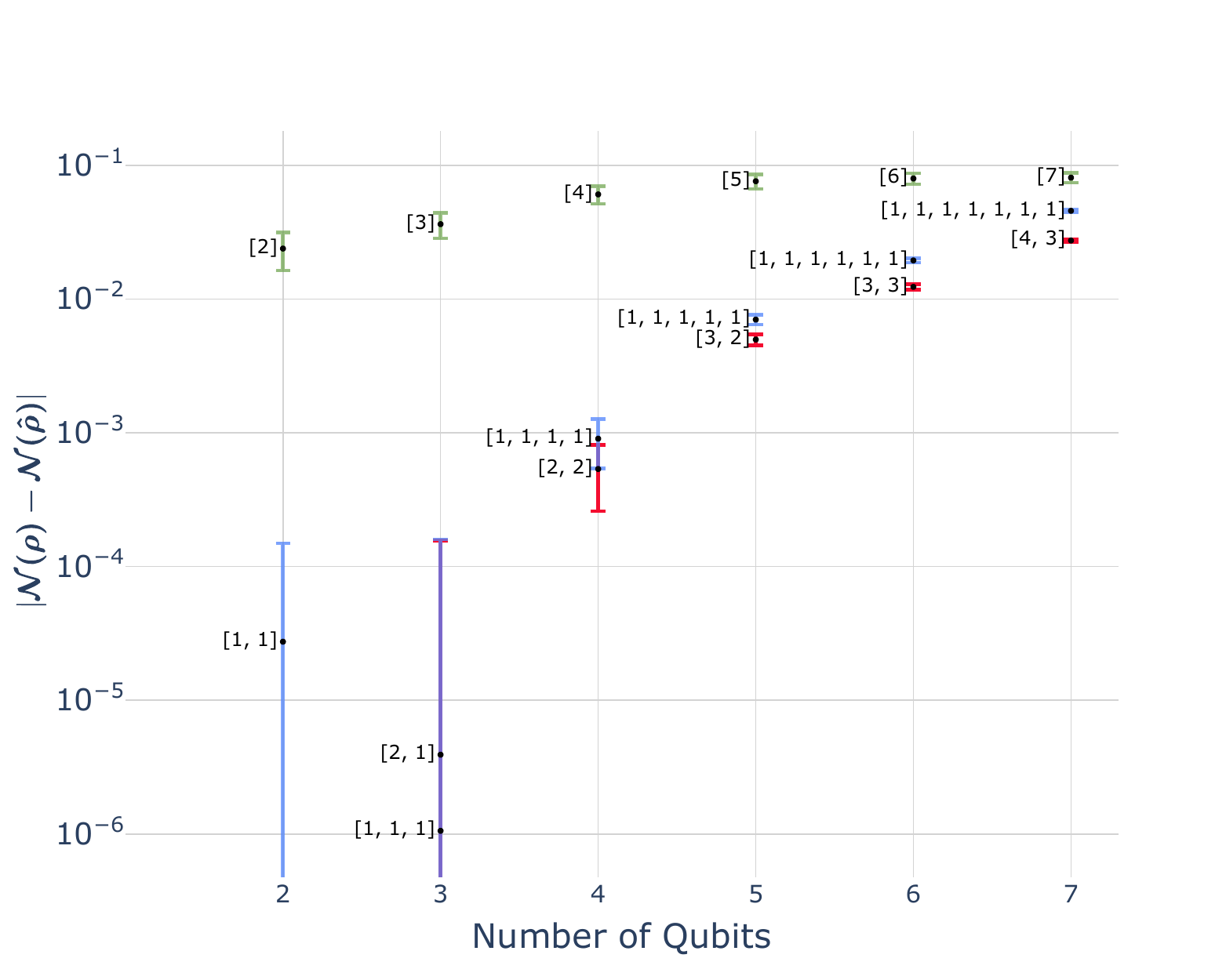}
    \caption{ Error on estimating entanglement negativity from the PLS for a sample size of $N=1.5\times10^7$. The labels next to each data point show the distribution of qubits among quantum nodes. Each data point corresponds to the average error from tomography of a 100 locally-random GHZ state. Error bars correspond to the standard deviation. }\label{fig:results_noise_negativity}
\end{figure*}

We also employ these noisy numerical simulations to study the use of the PLS estimator to estimate the entanglement negativity between the two quantum nodes. From Fig.\thinspace\ref{fig:results_noise_negativity}, we see that, as in the trace-norm error case, tomography with node 2-designs provides the best estimate of entanglement negativity for a large number of qubits. For a small number of qubits, tomography using local Pauli measurements yields errors comparable to those from node 2-designs, particularly when accounting for error bars. Negativity from the global 2-design is stuck at an error of $0.1$ due to noise in the quantum link.

\section{Conclusions}

We introduced a PLS quantum state tomography tailored for distributed quantum computers that avoids assuming remote entanglement as a primitive resource. By constructing separable POVMs from node projective 2-designs, the protocol enables the characterization of multipartite quantum states using only trusted entangling operations within each node and classical communication between nodes. We demonstrated that to obtain an estimate with a trace-norm error $\epsilon$, our protocol requires a sample of size $\mathcal{O}(2^Mr^2d\log(d)/\epsilon^2)$. This result includes previously known results in tomography via Pauli measurements and global 2-designs. The exponential dependence on the number of subsystems reflects the statistical cost of performing tomography using local 2-design measurements. Additionally, we consider the problem of estimating remote entanglement using the PLS tomographic estimator and establish certified error bounds for the entanglement Negativity. Numerical simulations using mutually unbiased bases support the theoretical analysis and demonstrate the expected degradation in reconstruction quality in distributed systems compared to individual devices with the same total number of qubits. 

The framework presented here establishes a principled baseline for benchmarking small-scale fault-tolerant distributed quantum computers (FT-DQC) \cite{Katabarwa2024_EFT,Larasati2025} and for certifying the generation of remote entanglement. A natural strategy for characterizing such devices is full-local Pauli tomography. However, our results show that this approach is not the most statistically efficient. Instead, better performance is achieved by minimizing the number of subsystems $M$, thereby exploiting the trusted entanglement available within each quantum node. In this regime, the protocol approaches global 2-design measurements, providing the optimal achievable reconstruction accuracy \cite{optimal_sampling_7956181}.

Our protocol is also directly applicable to near-term quantum devices \cite{Preskill2018quantumcomputingin}. In this regime, noisy operations and limited coherence times restrict implementations to shallow circuits, which, in general, prevent the realization of large node 2-designs. To overcome this limitation, a single physical device can be partitioned into multiple virtual nodes, each containing a sufficiently small number of qubits to reliably implement the corresponding node 2-design. The size of these virtual nodes can be estimated, for instance, via the quantum volume \cite{PhysRevA.100.032328}. A device with quantum volume $2^k$ can implement random circuits of width and depth $k$ with high success probability. Since the circuits for MUBs require $\mathcal{O}(n^2)$ gates, this indicates that reliable implementations are limited to subsystems of at most $\mathcal{O}(\sqrt{k})$ qubits. Alternative estimates can be obtained from coherence times, and two-qubit gate fidelities, and the use of error mitigation can increase the size of reliably implemented node 2-designs \cite{RevModPhys.95.045005}.

A key feature of our approach is that the number of subsystems $M$ can be treated as a tunable parameter. In particular, techniques based on dynamical circuits enable the effective redefinition of subsystem sizes through mid-circuit measurements and classical feedback \cite{PhysRevLett.127.100501}. An example of this is circuit knitting \cite{Knitting_10236453,https://doi.org/10.48550/arxiv.2603.12411}, a family of techniques that decompose large quantum circuits into smaller subcircuits that can be executed independently, with the global outcome reconstructed via classical post-processing. From this perspective, circuit knitting provides an operational tool for modifying the effective partitioning of a quantum system. In the FT-DQC regime, it enables combining multiple nodes into larger effective subsystems \cite{CarreraVazquez2024}, reducing $M$ and improving sample complexity. In near-term devices, it allows decomposing a system into smaller virtual nodes that are compatible with hardware constraints without increasing the number of subsystems \cite{10.1145/3676536.3676719}. However, note that not all quantum circuits can be efficiently decomposed via circuit knitting \cite{https://doi.org/10.48550/arxiv.2404.03619}. Consequently, the partitioning of MUB circuits requires careful optimization in terms of system size and hardware connectivity to ensure practical usability.

The exponential dependence of the sample complexity on the number of subsystems suggests that fully general tomography will be impractical for large-scale distributed networks. This motivates the development of more resource-efficient characterization methods that leverage the trusted entanglement. One way to avoid exponential scaling in tomography is to incorporate prior information about the unknown state, such as its purity \cite{PhysRevLett.115.090401, Carmeli2016, Pereira2022, Zambrano2024, 2510.16117}, rank \cite{PhysRevLett.105.150401, Flammia2012}, or small bond dimension \cite{Cramer2010, Lanyon2017, Torlai2023,rbg2-f61m}. Formulating distributed versions of these protocols could permit efficient full characterization of large distributed quantum computers through tomography. Beyond full device characterization, partial characterization using non-overlapping tomography \cite{PhysRevLett.124.100401, PhysRevA.106.062441, Pereira2023} appears well-suited for distributed quantum computers. In this protocol, a set of reduced density matrices is estimated, which can be done efficiently if we restrict to the device connectivity. Similarly, we can characterize a distributed quantum device by a set of reduced density matrices that encode the quality of all quantum links. Remote entanglement can also be directly estimated by collective measurements \cite{PhysRevLett.89.127902, PhysRevLett.90.167901, PhysRevLett.102.190503}, variational quantum algorithms \cite{Wang2022,PhysRevApplied.18.024048,Zambrano2024_GME_BP,CortsVega2023,839s-4qqv} or classical shadows \cite{Huang2020,10.1145/3188745.3188802, PhysRevLett.125.200501, PhysRevX.14.031035}, without resorting to tomography. Designing distributed versions of these protocols could provide an efficient strategy for estimating or certifying remote entanglement in large distributed architectures.

\section*{Acknowledgement}
L.P. was supported by the Government of Spain (Severo Ochoa CEX2019-000910-S, FUNQIP, and QEC4QEA PCI2025-163167), European Union (PASQuanS2.1, 101113690 and QEC4QEA, 101194322), Fundació Cellex, Fundació Mir-Puig, and Generalitat de Catalunya (CERCA program). A.D. and H.M.-V. were supported by the National Agency of Research and Development (ANID) -- Millennium Science Initiative Program -- ICN17$_-$012. A.D. acknowledges financial support from FONDECYT Regular Grant No. 1230586.

\newpage 
\appendix 

\section{Least-squares estimator}\label{ap:SLE}

In this section, we show that the solution of the least-squares problem defined in Eq.\thinspace\eqref{eq:LSP} coincides with the estimator given in Eq.\thinspace\eqref{eq:LSE}.

We begin by recalling some results on least-squares tomography with global 2-designs. Let $\{ \dyad{v_k} \}$ be a $m$-outcomes projective 2-design on dimension $d$, and consider the linear map $\mathcal{M}(X)_k=d\Tr( \dyad{v_k} X )/m$. In \cite{Guta2020} it was shown that the map $\mathcal{M}$ satisfies
\begin{align}
    (\mathcal{M}^\dagger\mathcal{M})^{-1}(X) = \frac{m}{d}\left[ (d+1)X -\Tr(X) \mathbb{I} \right], \label{eq:MdagM_global}
\end{align}
and that the solution of the least-squares problem  is
\begin{align}
    \hat{L} = (d+1) \sum_k f_k \left[ \dyad{v_k} - \frac{\mathbb{I}}{d+1} \right]. \label{eq:L_global}
\end{align}
These results can be used to calculate the LS estimator for the POVM defined by node 2-designs of Eq.\thinspace\eqref{eq:POVM_local}. 

Let us consider $M$ quantum systems of dimensions $d_j$, so the total dimension is $d=d_1\cdots d_M$. Let us consider $X\in\mathbb{H}^{d_1}\otimes\cdots\otimes\mathbb{H}^{d_M}\equiv\mathbb{H}^{d}$ a Hermitian matrix. It can be expanded in a local operator basis as
\begin{align}
    X = \sum_{k_1\cdots k_M} c_{k_1\cdots k_M} X_{k_1}\otimes\cdots\otimes X_{k_M}.
\end{align}
Consider the linear map $\mathcal{M}$ of Eq.\eqref{eq:linear_map} defined by node 2-designs of Eq.\eqref{eq:POVM_local}. Since $\mathcal{M}$ is linear, the operator $(\mathcal{M}^\dagger\mathcal{M})^{-1}$ is also linear. Thus
\begin{align}
    (\mathcal{M}^\dagger\mathcal{M})^{-1}(X) = \sum_{k_1\cdots k_M} c_{k_1\cdots k_M} (\mathcal{M}^\dagger\mathcal{M})^{-1}( X_{k_1}\otimes\cdots\otimes X_{k_M} ).
\end{align}
Therefore, it suffices to analyze the action of $(\mathcal{M}^\dagger\mathcal{M})^{-1}$ on a tensor-product operator $\bigotimes_{j=1}^MX_{j}$. Noticing that
\begin{align}
    \mathcal{M}\left(\bigotimes_{j=1}^MX_{j}\right)_{k_1\cdots k_M} = \frac{d}{m}\Tr\!\left( \bigotimes_{j=1}^M \dyad*{v_{k_j}^j} X_{j} \right), 
\end{align}
the map factorizes into a product of traces over the individual subsystems,
\begin{align}
    \mathcal{M}\left(\bigotimes_{j=1}^MX_{j}\right)_{k_1\cdots k_M} = \frac{d}{m}\prod_{j=1}^M\Tr\!\left( \dyad*{v_{k_j}^j} X_{j} \right)=\prod_{j=1}^M \mathcal{M}_j(X_{j})_{k_j},
\end{align}
with $\mathcal{M}_j$ the map restricted to the $j$-th subsystem. Recalling that
\begin{align}
    \mathcal{M}^\dagger\mathcal{M}\left(X\right)=  \sum_{k_1\cdots k_j}^{m_1\cdots m_M} \mathcal{M}\left(X\right)_{k_1\cdots k_M} \bigotimes_{j=1}^M\dyad*{v_{k_j}^j},
\end{align}
from which it follows that
\begin{align}
    \mathcal{M}^\dagger\mathcal{M}\left(\bigotimes_{j=1}^MX_{j}\right) = \bigotimes_{j=1}^M \sum_{k_j=1}^{m_j} \mathcal{M}_j(X_{j})_{k_j}\dyad*{v_{k_j}^j} = \bigotimes_{j=1}^M \mathcal{M}_j^\dagger\mathcal{M}_j\left(X_{j}\right).
\end{align}
Therefore
\begin{align}
(\mathcal{M}^\dagger\mathcal{M} )^{-1}\left( \bigotimes_{j=1}^MX_{j}\right) = \bigotimes_{j=1}^M ( \mathcal{M}_j^\dagger\mathcal{M}_j )^{-1} \left(X_{j}\right).
\end{align}
Using Eq.\thinspace\eqref{eq:MdagM_global}, we obtain
\begin{align}
    (\mathcal{M}^\dagger\mathcal{M} )^{-1}\left( \bigotimes_{j=1}^MX_{j}\right) = \bigotimes_{j=1}^M \frac{m_j}{d_j} \left[ (d_j+1)X_j -\Tr(X_j) \mathbb{I}_j \right] = \frac{m}{d} \bigotimes_{j=1}^M \left[ (d_j+1)X_j -\Tr(X_j) \mathbb{I}_j \right]. \label{eq:MdagM_inv_local}
\end{align}
Now we can calculate the least-squares estimator
\begin{align}
    \hat{L} = (\mathcal{M}^\dagger\mathcal{M} )^{-1}\mathcal{M}^\dagger( f_{k_1\cdots k_j} ) = (\mathcal{M}^\dagger\mathcal{M} )^{-1} \left( \frac{d}{m} \sum_{k_1\cdots k_M}^{m_1\cdots m_M}f_{k_1\cdots k_M} \bigotimes_{j=1}^M \dyad*{v_{k_j}^j} \right).
\end{align}
Using linearity, we have
\begin{align}
    \hat{L} = \frac{d}{m} \sum_{k_1\cdots k_M}^{m_1\cdots m_M}f_{k_1\cdots k_M}  (\mathcal{M}^\dagger\mathcal{M} )^{-1} \left( \bigotimes_{j=1}^M \dyad*{v_{k_j}^j} \right).
\end{align}
Employing Eq.\thinspace\eqref{eq:MdagM_inv_local}, we obtain
\begin{align}
    \hat{L} = \frac{d}{m} \sum_{k_1\cdots k_M}^{m_1\cdots m_M}f_{k_1\cdots k_M} \frac{m}{d} \bigotimes_{j=1}^M \left[ (d_j+1)\dyad*{v_{k_j}^j} -\Tr(\dyad*{v_{k_j}^j}) \mathbb{I}_j \right]. 
\end{align}
Expanding this expression yields Eq.\thinspace\eqref{eq:LSE} of the main text.

\section{Concentration inequality}\label{ap:conc_ineq}

In this section, we derive rigorous non-asymptotic error bounds for the PLS estimator. We first derive an operator-norm concentration bound for the LS estimator and then convert it into a trace-norm bound for the PLS estimator. Our analysis is based on the matrix Bernstein inequality introduced in \cite{Tropp2011}: 
\begin{theorem}\label{theo:bernstein}
Let ${A_k}\in\mathbb{H}$ be a sequence of independent Hermitian random matrices satisfying
\begin{align}
\mathbb{E}[A_k] = 0, \qquad \norm{A_k}_\infty \leq R \quad \text{almost surely}. \label{eq:bound_norm}
\end{align}
Then, for any $\epsilon > 0$,
\begin{align} 
\mathbb{P}\left[ \norm{\sum_{a=1}^NA_{k^a}}_\infty \geq \epsilon \right] \leq \Bigg\{ \begin{matrix} d \exp\left(-\frac{3\epsilon^2}{8\sigma^2} \right), \quad \epsilon\leq \frac{\sigma^2}{R},\\ d \exp\left(-\frac{3\epsilon}{8R} \right), \quad \epsilon > \frac{\sigma^2}{R}, 
\end{matrix} 
\end{align}
where
\begin{align}
\sigma^2 = \norm{ \sum_{a=1}^N \mathbb{E}[A_{k^a}^2] }_\infty . \label{eq:bound_variance}
\end{align}
\end{theorem}
Here, $\norm{A}_\infty$ is the operator norm, defined as the largest singular value of $A$, and the index $k^a=(k_1^a,\cdots,k_M^a)$ represents the outcome obtained in the $a$-th single-shot measurement. We now apply Theorem~\ref{theo:bernstein} to the least-squares estimator $\hat L$ given by our tomography protocol. Note that $\hat L$ can be written as the average of random matrices associated with single-shot measurement outcomes,
\begin{align}
    \hat L = \frac{1}{N}\sum_{a=1}^N X_{k^a}
\end{align}
where we define the random matrices $X_{k}$ as 
\begin{align} 
X_{k} = X_{k_1\cdots k_M} = \bigotimes_{j=1}^M(d_j+1)\left[\dyad*{v^j_{k_j}}{v^j_{k_j}}-\frac{\mathbb{I}_j}{d_j+1}\right]. \label{eq:random_matrices}
\end{align}
By construction, the expectation value of $X_k$ satisfies $\mathbb{E}[X_k] = \rho$. We therefore consider the centered random matrices $A_k = (\rho - X_k)/N$, which satisfy $\mathbb{E}[A_k]=0$. In order to apply the theorem \ref{theo:bernstein} to $A_k$, we must find the parameters defined by Eqs. \eqref{eq:bound_norm} and \eqref{eq:bound_variance}.

We first compute the parameter $R$. Given that $A_j$ has positive and negative eigenvalues, the spectral norm can be expressed as 
\begin{align}
    \norm{A_j}_\infty =\max( \lambda_j^{\rm max}, -\lambda_j^{\rm \min} ) ,
\end{align}
with $\lambda_j^{\rm max}$ and $\lambda_j^{\rm min}$ the maximum and minimum eigenvalues of $A_j$, respectively. Then, in order to upper bound the spectral norm we have to find an upper bound for the largest eigenvalue and a lower bound for the smallest. Let us consider an arbitrary state $\ket{\psi}$. Thus
\begin{align}
    \ev{A_j}{ \psi} = \frac{1}{N}\left[ \ev*{\bigotimes_{j=1}^N \left[(d_j+1)\dyad*{v_{k_j}^j} -\mathbb{I}_j\right]}{ \psi}  - \ev{\rho}{ \psi} \right]. 
\end{align}
This expected value can be upper- and lower-bounded as
\begin{align}
    \ev{A_j}{ \psi} & \leq \frac{1}{N} \ev*{\bigotimes_{j=1}^N \left[(d_j+1)\dyad*{v_{k_j}^j} -\mathbb{I}_j\right]}{ \psi} \leq \frac{1}{N} \prod_{j=1}^N d_j = \frac{d}{N},\\
    \ev{A_j}{ \psi} & \geq  \frac{1}{N}\left[ -\ev*{\bigotimes_{j=1}^N \mathbb{I}_j}{ \psi}  - \ev{\rho}{ \psi} \right] \geq \frac{-2}{N}. 
\end{align}
Since $\ket{\psi}$ was an arbitrary state and the bounds are independent of $\ket{\psi}$, we have 
\begin{align}
    \lambda_j^{\rm max} \leq \frac{d_j}{N}, \qquad \lambda_j^{\rm \min}\geq -\frac{2}{N}.
\end{align}
Therefore, the spectral norm of $A_j$ can be upper-bounded as
\begin{align}
    \norm{A_j}_\infty \leq \frac{1}{N}\max( d, 2 ) = \frac{d}{N}\equiv R.
\end{align}

For calculating the constant $\sigma^2$, notice that
\begin{align}
    \mathbb{E}\left[A_{k}^2 \right] = \frac{1}{N}\mathbb{E}\left[ X_k^2 -2X_k\rho + \rho^2 \right] =\frac{1}{N}\left( \mathbb{E}[X_k^2] - \rho^2 \right),
\end{align}
where we used that $\mathbb{E}[X_k]=\rho$. Thereby,
\begin{align}
    \norm{\frac{1}{N}\sum_{a=1}^N\mathbb{E}\left[A_{k^a}^2 \right]}_\infty \leq \frac{1}{N^2} \sum_{a=1}^N\norm{ \mathbb{E}[X_{k^a}^2]-\rho^2  }_\infty
\end{align}
Then, we must calculate
\begin{align}
    \mathbb{E}[X_{k}^2] = \mathbb{E}\left[ \bigotimes_{j=1}^M \left[ (d_j+1)\dyad*{v_{k_j}^j} - \mathbb{I}_j \right]^2 \right] = \bigotimes_{j=1}^M \left[ (d_j-1)(\mathbb{I}_j+\rho_j) + \mathbb{I}_j \right],
\end{align}
where $\rho_j$ is the reduced density matrix of $\rho$ in the $j$-th subsystem, that is,
\begin{align}
    \rho_j = \Tr_{[1,\cdots,j-1,j+1,\cdots,M]}(\rho).
\end{align}
Similarly to the calculation of $R$, to bound $\sigma$ we have to bound the largest and smallest eigenvalues of $\mathbb{E}[X_{k}^2]-\rho^2$. Considering an arbitrary state $\ket{\psi}$, we have
\begin{align}
    -1 \leq \ev{ \left( \mathbb{E}[X_{k}^2]-\rho^2 \right) }{\psi} \leq \prod_{j=1}^M (2d_j-1).
\end{align}
Therefore,
\begin{align}
    \norm{\frac{1}{N}\sum_{a=1}^N\mathbb{E}\left[A_{k^a}^2 \right]}_\infty \leq \frac{1}{N^2} \sum_{a=1}^N\max\left(  \prod_{j=1}^M (2d_j-1), 1\right)  \leq \frac{1}{N} \prod_{j=1}^M (2d_j-1).
\end{align}
Therefore, we can set
\begin{align}
    \sigma^2 = \frac{2^M}{N}\prod_{j=1}^M(d_j-1/2).
\end{align}
Noting that we can express $\sigma^2$ in a more elegant form, with an upper bound of $\sigma^2 \leq 2^M d/N$. However, we prefer to use the original expression as it is more tightly bound for small dimensions and aligns with previous results regarding Pauli measurements \cite{Guta2020}. This scenario is obtained by setting $d_j = 2$, in which case the 2-designs reduce to Pauli measurements in each subsystem, yielding $\sigma^2 = 3^M/N$. 

Applying Theorem~\ref{theo:bernstein} to the sum $\sum_{a=1}^N A_{k^a} = \rho - \hat L$, we obtain the following concentration inequality for the operator-norm error of the least-squares estimator,
\begin{align} 
\mathbb{P}\left[ \norm*{ \rho - \hat{L} }_\infty \geq \epsilon \right] \leq d \exp\left( -\frac{3\epsilon^2N}{8\times2^M\prod_{j=1}^M(d_j-1/2)} \right), \label{eq:conc_inq_infty_norm} 
\end{align}
valid for $\epsilon \leq 2^M$. This concentration inequality in terms of the operator norm can be converted into the trace norm $\norm{A}_1=\Tr\sqrt{A^\dagger A}$, considering the following proposition introduced in \cite{Guta2020}:
\begin{proposition}\label{theo:proposition}
    Let $\hat{L}\in\mathbb{H}_d$ be the LS estimator of a state $\rho\in\mathbb{H}_d$. Then, for any $r\in\mathbb{N}$, the PLS estimator $\hat{\rho}$ satisfies
    \begin{align}
        \norm{\rho-\hat\rho}_1 \leq 4r\norm*{\rho-\hat{L}}_\infty + 2\min\{ \Lambda_r(\rho), \Lambda_r(\hat\rho) \},
    \end{align}
    with $\Lambda_r(\rho) =\min_{{\rm rank}(Z)\leq r} \norm{\rho-Z}_1 $.
\end{proposition}
In simple words, this proposition tells us that the trace-norm distance between $\rho$ and $\hat{\rho}$ is upper-bounded by $4r$ times the operator-norm distance between $\rho$ and $\hat{L}$ plus a bias depending on $\Lambda_r(\rho)$. This quantity is defined as an optimization problem, whose solution $\hat Z$ corresponds to the truncation of $\rho$ to the largest contributions $r$ in the decomposition of the eigenvalue. Thus, $\Lambda_r(\rho)$ is the residual error for truncation $\rho$ up to rank-$r$. For simplicity, we will consider $r=\min\{ {\rm rank}(\rho), {\rm rank}(\hat\rho) \}$, so that the truncation error vanishes. Although the rank of $\rho$ is generally unknown, this choice highlights the intrinsic low-rank scaling of the bound. In practice, $r$ may be chosen as an effective rank or via a noise model. By combining Proposition \ref{theo:proposition} with Eq.\thinspace\eqref{eq:conc_inq_infty_norm}, we have
\begin{align}
    \mathbb{P}\left[ \norm*{ \rho - \hat{L} }_1 \geq 4r\epsilon \right] \leq d \exp\left( -\frac{3\epsilon^2N}{8\times2^M\prod_{j=1}^M(d_j-1/2)} \right). 
\end{align}
Redefining the error bound $\epsilon\rightarrow\epsilon/4r$, we obtain
\begin{align}
    \mathbb{P}\left[ \norm*{ \rho - \hat{L} }_1 \geq \epsilon \right] \leq d \exp\left( -\frac{3(\epsilon/4r)^2N}{8\times2^M\prod_{j=1}^M(d_j-1/2)} \right), 
\end{align}
which allows us to formulate Theorem \ref{theo:main} in the main text.

\newpage
\section{More Numerical Simulations}\label{apex:sims}

In this appendix, we present some additional numerical simulations that complement the results presented in the main text.

\subsection{Simulation for other sample sizes}
\begin{figure*}[h!]
    \centering
    \begin{subfigure}[t]{0.53\textwidth}
        \centering
        \includegraphics[width=\linewidth]{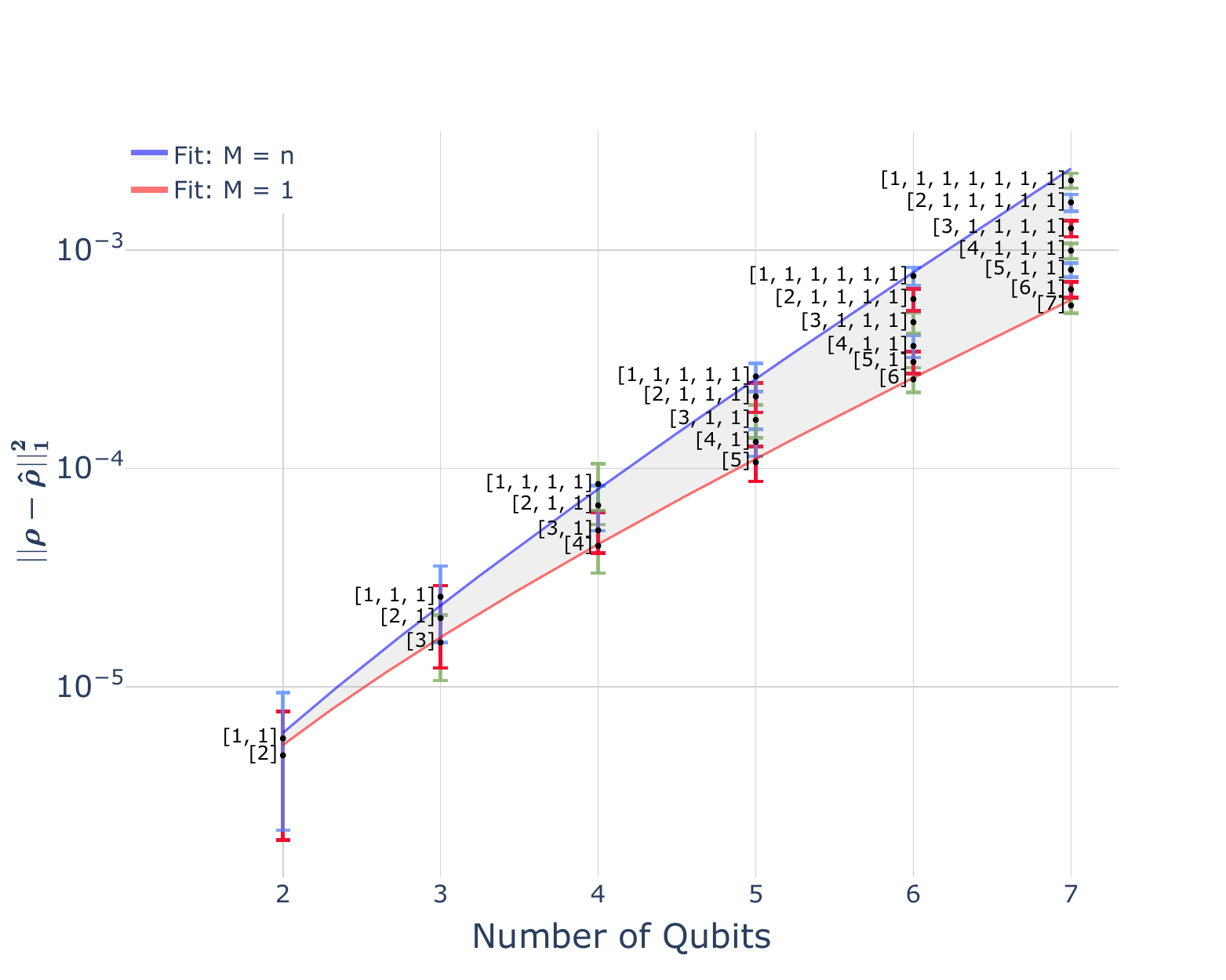}
        \caption{}
    \end{subfigure}%
    \begin{subfigure}[t]{0.53\textwidth}
        \centering
        \includegraphics[width=\linewidth]{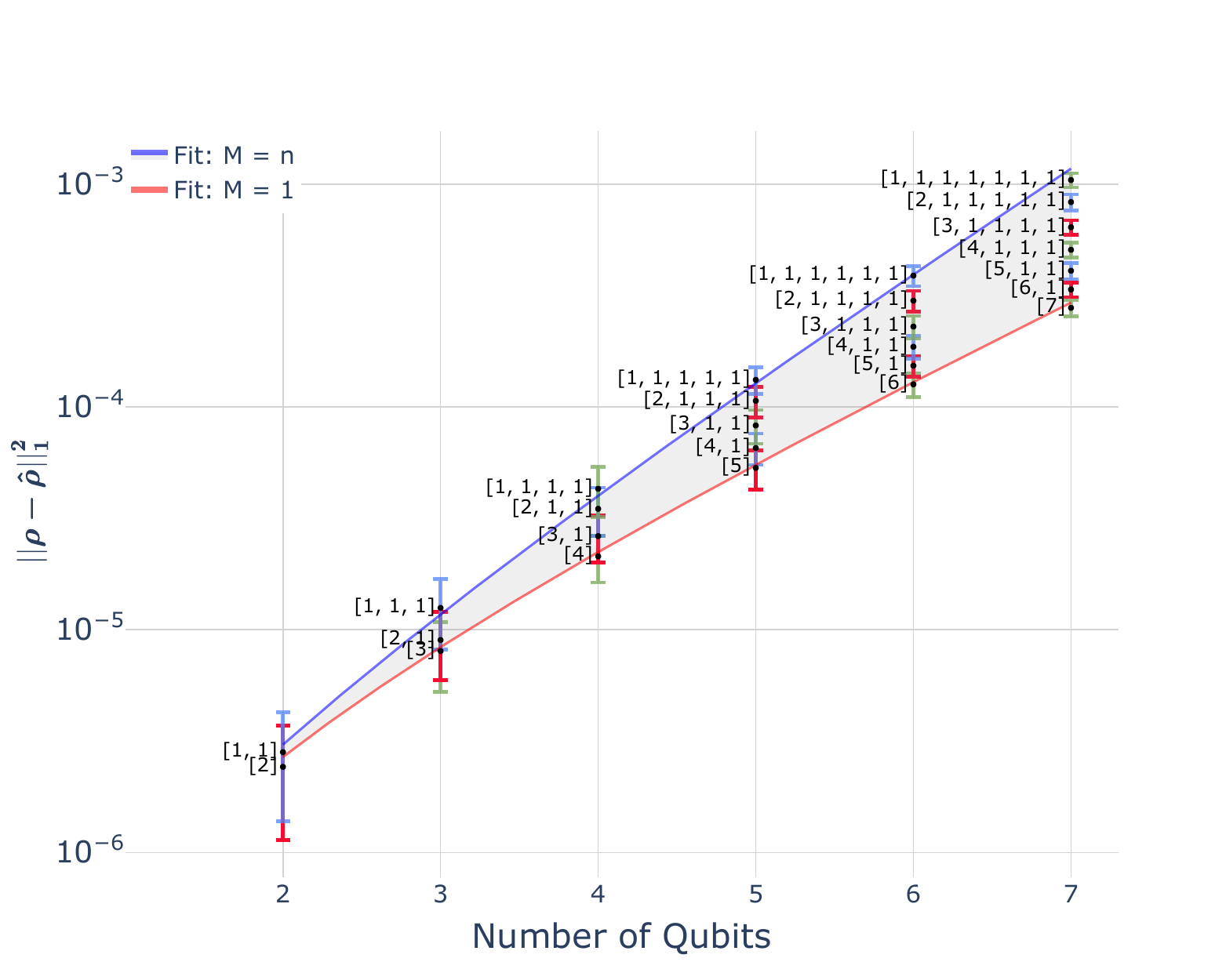}
        \caption{}
    \end{subfigure}
    \caption{ Numerical simulation of PLS tomography with sample sizes (a) $N=5\times10^6$ and (b) $N=10^7.$ The labels next to each data point show the distribution of qubits among quantum nodes. Each point corresponds to the average trace-norm error from the tomography of 100 random Haar states, while the error bars correspond to their standard deviation. The solid lines correspond to the heuristic scaling of Eq.\thinspace\eqref{eq:heuristic_scaling} for $M = 1$ (red) and $M = n$ (blue). }\label{fig:results_apx}
\end{figure*}

Figure \ref{fig:results_apx} displays the numerical simulation for sample sizes $N=5\times10^6$ and $N=10^7$. We observe the same overall behavior as Fig.\thinspace\ref{fig:results} of the main text.

\newpage
\subsection{Simulation for other depolarization noises}
\begin{figure*}[h!]
    \centering
    \begin{subfigure}[t]{0.53\textwidth}
        \centering
        \includegraphics[width=\linewidth]{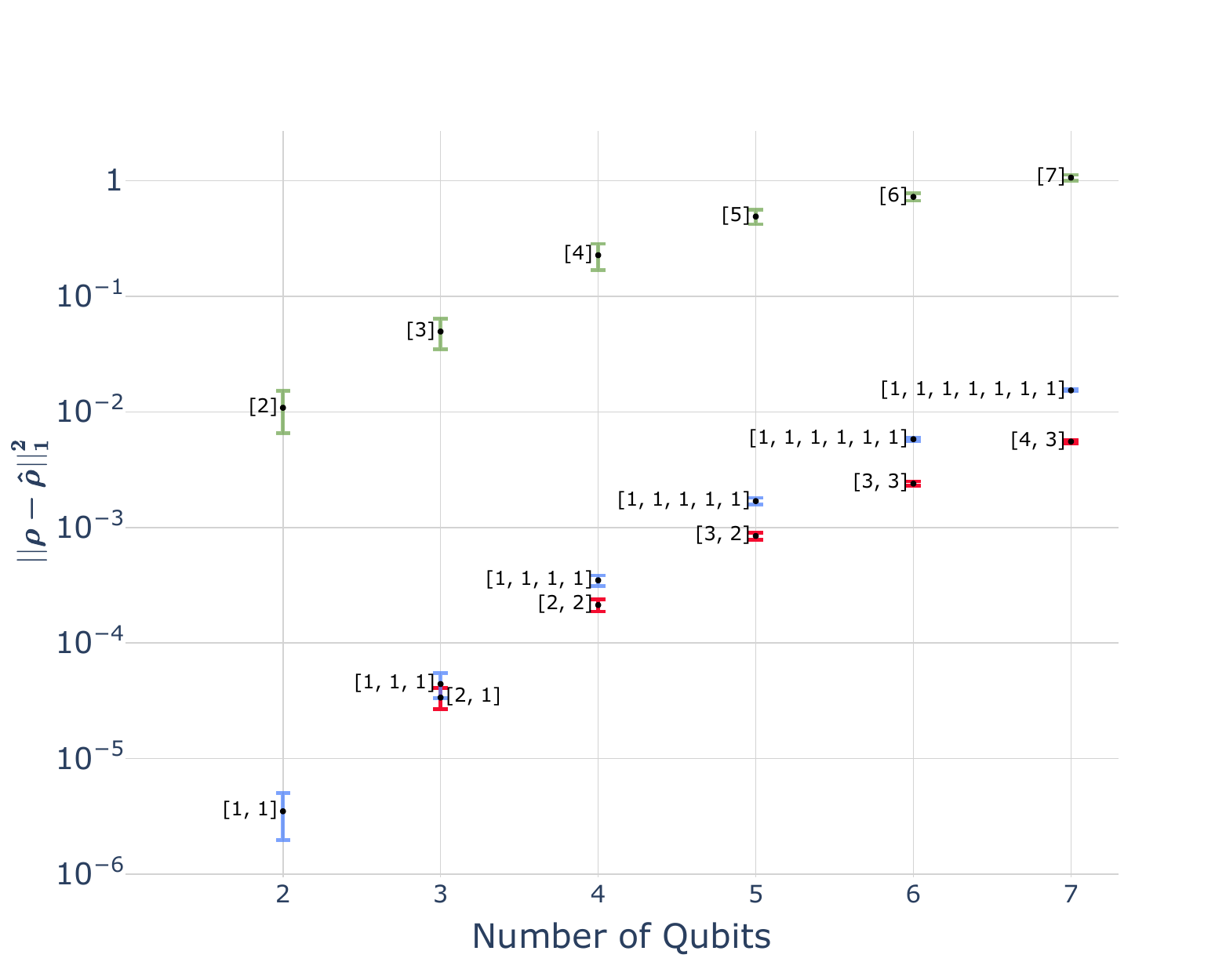}
        \caption{}
    \end{subfigure}%
    \begin{subfigure}[t]{0.53\textwidth}
        \centering
        \includegraphics[width=\linewidth]{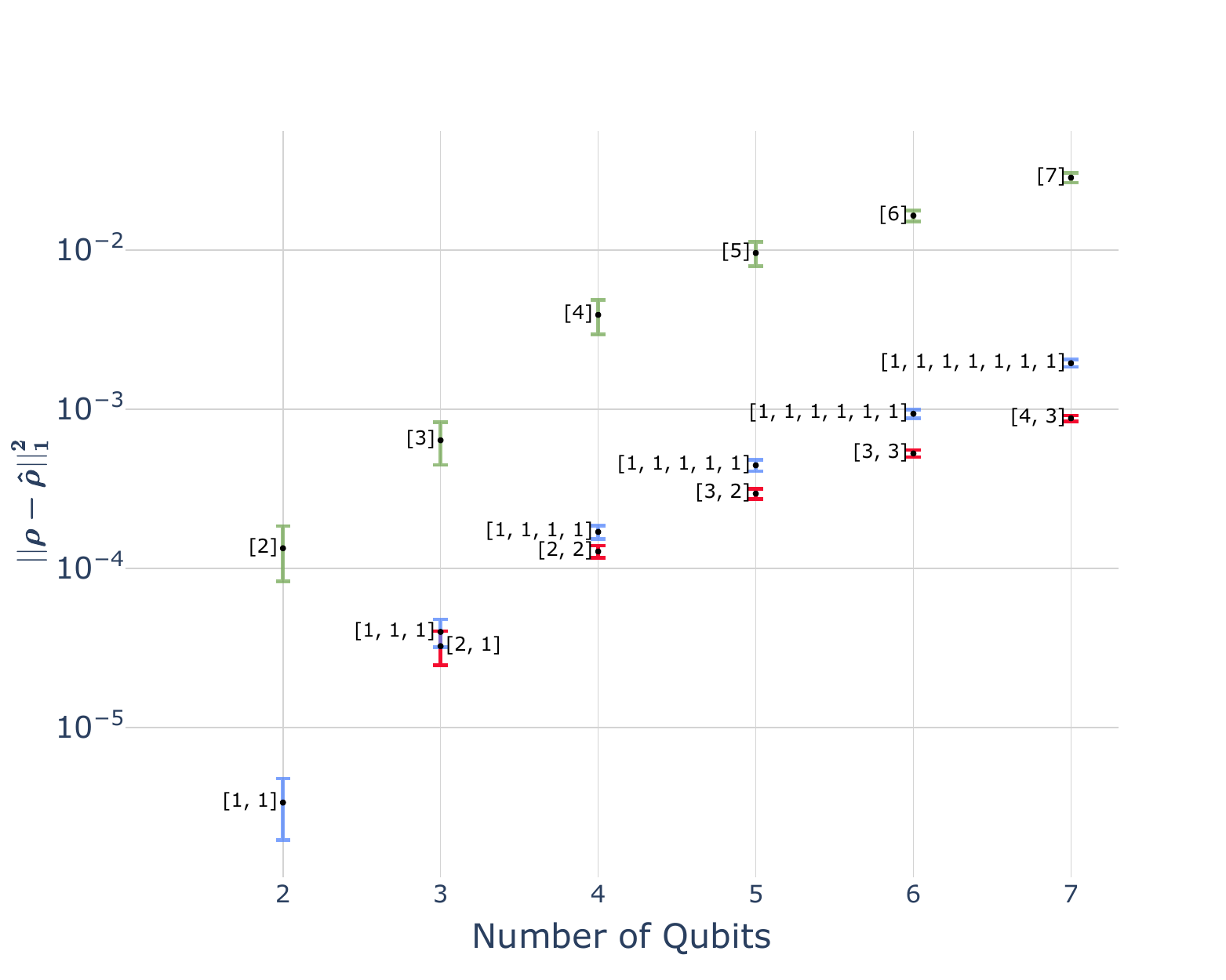}
        \caption{}
    \end{subfigure}
    \caption{ Numerical simulation of PLS tomography for depolarization noises (a) $\lambda=0.1$ and (b) $\lambda=0.01$. The labels next to each data point show the distribution of qubits among quantum nodes. Each point corresponds to the average trace-norm error from the tomography of a 100 locally-random GHZ state, while the error bars correspond to their standard deviation. }\label{fig:results_apx_noise}
\end{figure*}

Figure \ref{fig:results_apx_noise} displays the numerical simulation of the tomography with Pauli measurements (blue), global 2-designs (green), and node 2-designs (red) for depolarization noises $\lambda=0.1$ and $\lambda=0.01$. We observe the same overall behavior as Fig.\thinspace\ref{fig:results_noise} of the main text. 

\begin{figure*}[h!]
    \centering
    \begin{subfigure}[t]{0.53\textwidth}
        \centering
        \includegraphics[width=\linewidth]{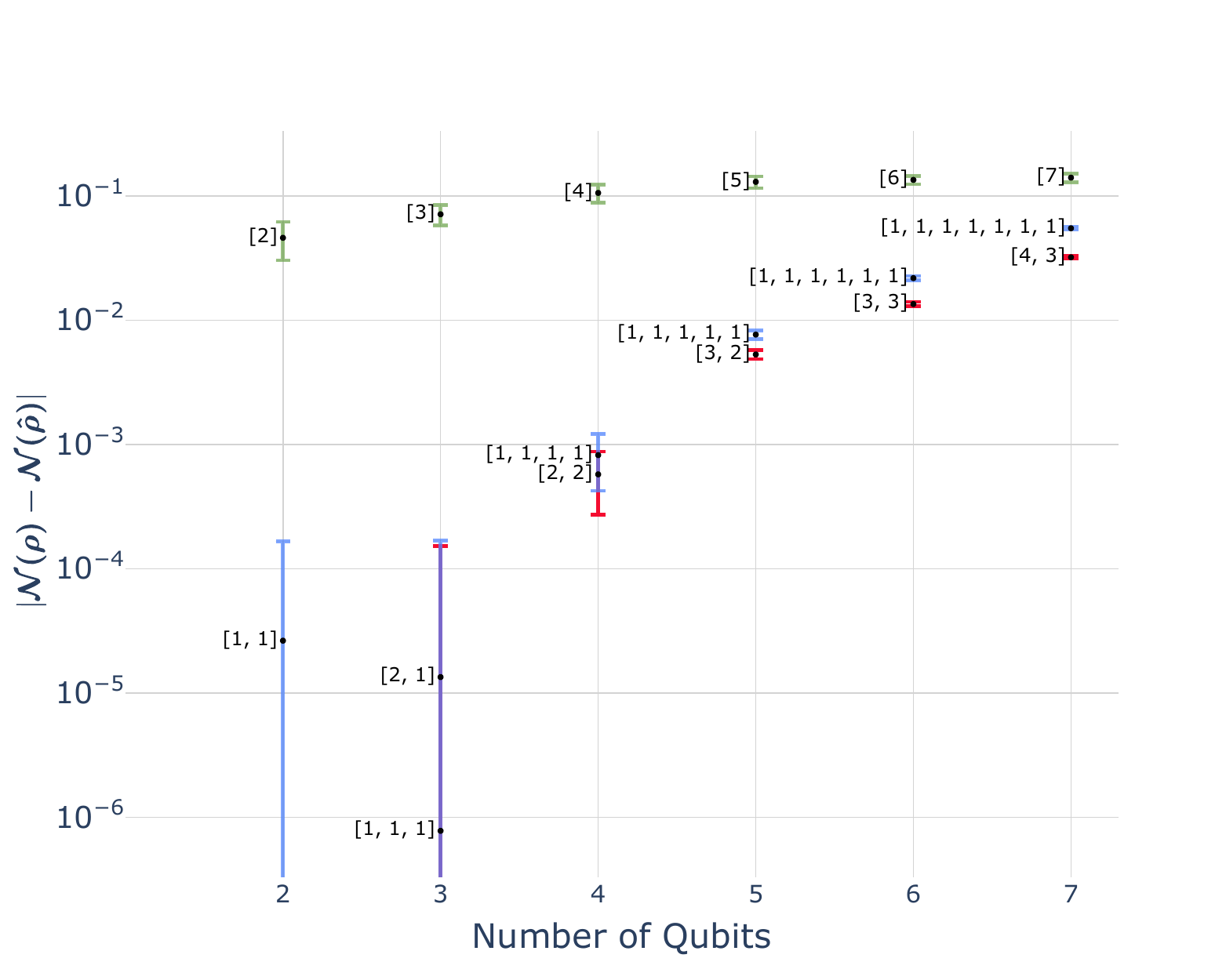}
        \caption{}
    \end{subfigure}%
    \begin{subfigure}[t]{0.53\textwidth}
        \centering
        \includegraphics[width=\linewidth]{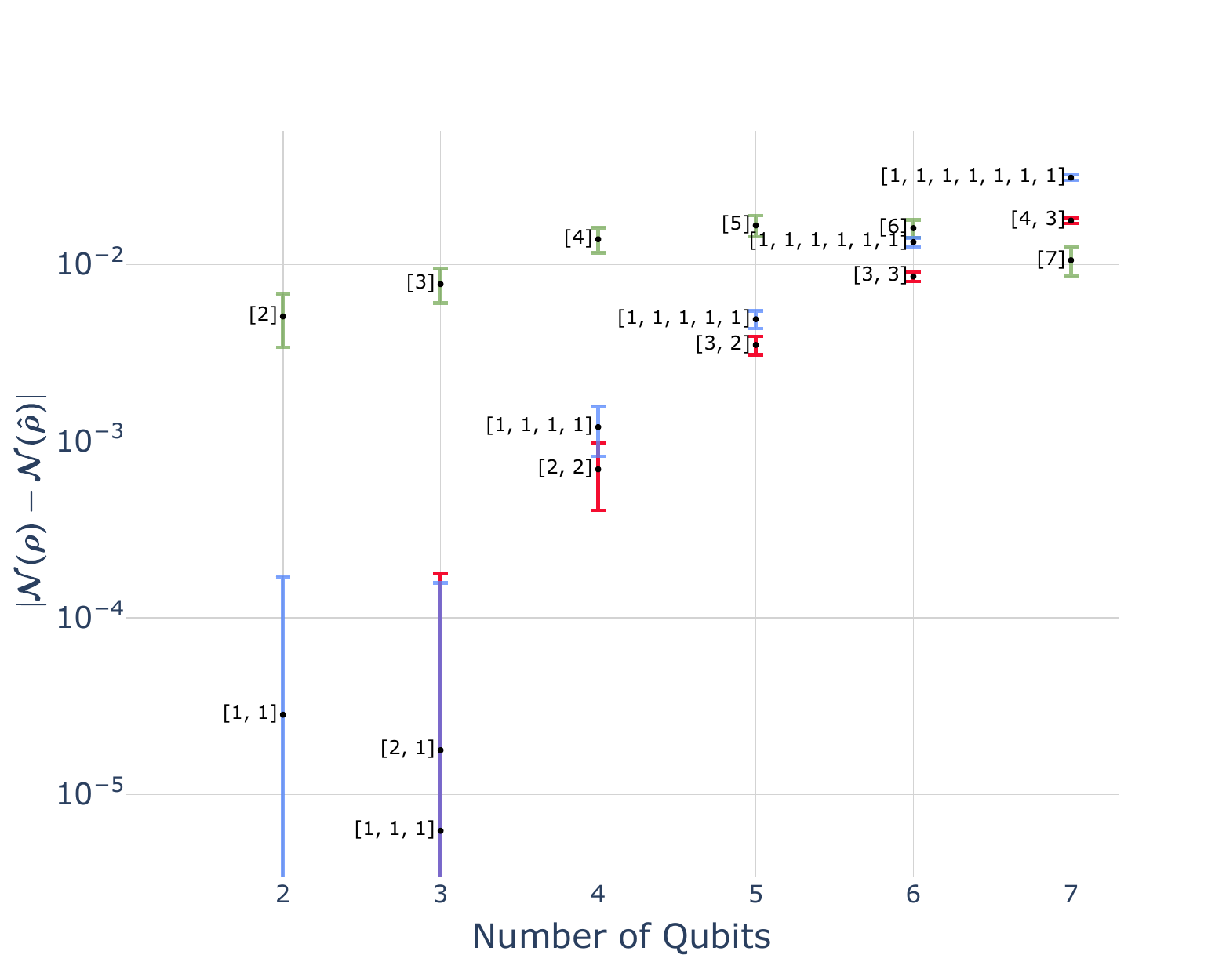}
        \caption{}
    \end{subfigure}
    \caption{ Numerical simulation of entanglement Negativity estimation from PLS tomography for depolarization noises $\lambda=0.1$ and $\lambda=0.01$. Error on estimating entanglement negativity from the PLS for a sample size of $N=1.5\times10^7$. The labels next to each data point show the distribution of qubits among quantum nodes. Each data point corresponds to the average error from tomography of a 100 locally-random GHZ state. Error bars correspond to the standard deviation.  }\label{fig:results_apx_noise_negativity}
\end{figure*}

Figure \ref{fig:results_apx_noise_negativity} displays the error on estimating the entanglement Negativity with tomography by Pauli measurements (blue), global 2-designs (green), and node 2-designs (red) for depolarization noises $\lambda=0.1$ and $\lambda=0.01$. We observe the same overall behavior as Fig.\thinspace\ref{fig:results_noise_negativity} of the main text. However, for 7 qubits with a parameter $\lambda = 0.01$, global 2-designs outperform local Pauli and node 2-designs in estimating the Negativity. This improvement is likely due to the effective depolarization noise on the Negativity being sufficiently low, making the statistical error the most significant factor.

\newpage
\subsection{Fitting constant}
We employ numerical simulation to obtain a heuristic scaling for the average trace-norm error. This is given by
\begin{align}
    \bar{\epsilon}  \approx \alpha \frac{(2^M)^{\beta} \tilde{d}^{\gamma} \log(d)}{N^\delta},
\end{align}
where the fitting constants, with their corresponding error, are
\begin{align} 
    \alpha =& 4.04 \pm 1.03,\\
    \beta  =& 0.795 \pm 0.015,\\
    \gamma =& 0.958 \pm 0.025,\\
    \delta =& 1.01 \pm 0.01.
\end{align}

\newpage
\section{Noisy GHZ state}\label{apx:GHZ}
An $n$-qubit Greenberger–Horne–Zeilinger (GHZ) state is given by
\begin{align}
    \ket{\Psi_{\rm GHZ}} = \frac{1}{\sqrt{2}}\left( \ket{0^{\otimes n}} + \ket{1^{\otimes n}} \right).
\end{align}
Suppose we prepare this state on a distributed quantum computer with two quantum nodes, each with $n_1$ and $n_2$ qubits. In addition, suppose that the remote CNOT between qubit $n_1$ and $n_1+1$ is affected by a depolarizing noise $\lambda$. Then, the density matrix of the noisy GHZ state is
\begin{align}
   \rho_{\rm GHZ} = (1-\lambda)\dyad{\Psi_{\rm GHZ}} + \frac{\lambda}{4} \mathbb{I}_{n_1,n_1+1} \Tr_{n_1,n_1+1}(\dyad{\Psi_{\rm GHZ}})
\end{align}
with $\mathbb{I}_{n_1,n_1+1}$ the identity matrix between the qubit $n_1$ and $n_1+1$ and $\Tr_{n_1,n_1+1}(\cdot)$ the partial trace in the qubits $n_1$ and $n_1+1$. Then, we have that
\begin{align}
    \rho_{\rm GHZ} = (1-\lambda)\dyad{\Psi_{\rm GHZ}} + \frac{\lambda}{8} \mathbb{I}_{n_1,n_1+1}\otimes\left( \dyad*{0^{\otimes n-2}} + \dyad*{1^{\otimes n-2}}\right).
\end{align}
We can easily see that this state has a rank of 8. 
\bibliographystyle{unsrt}
\bibliography{bib}

\end{document}